\documentclass[reprint,
 amsmath,amssymb,
 aps,floatfix,twocolumn, superscriptaddress]{revtex4-2}
\usepackage{times}
\usepackage{graphicx}
\usepackage{dcolumn}
\usepackage{bm}
\usepackage{amsmath}
\usepackage{braket}
\usepackage{xcolor}
\usepackage{dsfont}
\usepackage[T1]{fontenc}
\usepackage{comment}

\begin{document}
\title{Quantum Sensing with Bright Two-Mode Squeezed Light in a Distributed Network of Gyroscopes}
\author{Priyanka M. Kannath}
 \affiliation{Department of Physics, Indian Institute of Space Science and Technology\\
 Thiruvananthapuram, Kerala, 695547, India\\}
\author{Girish S. Agarwal}
 \affiliation{Institute for Quantum Science and Engineering, Department of Physics and Astronomy, Department of Biological and Agricultural Engineering\\ Texas A\&M University, College Station, Texas 77843, USA}
\author{Ashok Kumar}
 \email{ashokkumar@iist.ac.in}
  \affiliation{Department of Physics, Indian Institute of Space Science and Technology\\
 Thiruvananthapuram, Kerala, 695547, India\\}
 
\date{\today}
\begin{abstract}
 Recent developments in quantum technologies have enabled significant improvements in the precision of optical sensing systems. This work explores the integration of distributed quantum sensing (DQS) with optical gyroscopes to improve the estimation accuracy of angular velocity. Utilizing bright two-mode squeezed states (bTMSS), which offer high photon numbers and strong bipartite quantum correlations, we propose a novel configuration that leverages continuous-variable entanglement across multiple spatially separated optical gyroscopes. Unlike traditional quantum sensing that enhances a single sensor, our approach focuses on estimating a global phase shift—corresponding to the average angular rotation—across distributed optical gyroscopes with quantum-enhanced sensitivity. We analyze the phase sensitivities of different bTMSS configurations, including $M$ mode-entangled bTMSS and separable M-bTMSS, and evaluate their performance through the quantum Cramér-Rao bound. The analysis shows that, with $5\%$ photon loss in every channel in the system, the proposed scheme shows a sensitivity enhancement of $\sim 9.3$ dB beyond the shot-noise limit, with an initial squeezing of $\sim 9.8$ dB. The present scheme has potential applications in quantum-enhanced inertial navigation and precision metrology within emerging quantum networks.
\end{abstract}

\maketitle

\section{INTRODUCTION}
With the advent of quantum technologies, the sensing arena has significantly improved the precision with which measurements are made \cite{ giovannetti1, pirandola, degen, pezze, braun}. The quantum enhancements are reflected in a wide variety of applications like spectroscopy \cite{schmitt}, optomechanical sensors \cite{pooser1,xia}, super-resolution imaging \cite{nair,lupo}, plasmonic sensing \cite{Dowran}, detection of the magnetic field \cite{bonato}, remote sensing and navigation \cite{dowlingclock, cai}, to name a few. Quantum optical sensing, in particular, uses the nonclassical states of light and/or nonclassical detection techniques to improve the performance of various sensing applications \cite{caves, Yurke2, bondurant, walls, yuen, plick}. With coherent states of light, the maximum sensitivity of estimating a parameter is limited by the shot-noise limit (SNL) that scales to $1/\sqrt{N}$, with $N$ being the total number of photons probing the unknown parameter. This limit can however be surpassed by using quantum resources, which enables one to attain the ultimate Heisenberg limit (HL) in the sensitivity that scales to $1/N$ \cite{ Ou2, Woodworth, Holland, steuernagel}.

To date, entanglement and squeezing are the most substantially used quantum optical features for enhancing sensing performances \cite{hofmann2007high, afek, lawrie}.  Squeezed states were proved to be useful in optical interferometry in the pioneering theoretical proposal by Caves \cite{caves} in 1981, later experimentally demonstrated and implemented by many \cite{xiao, grangier}, a notable example being the Advanced Laser Interferometer Gravitational-Wave Observatory (LIGO) \cite{LIGO}. Among the quantum strategies utilizing the squeezed light, the bright two-mode squeezed states (bTMSS) offer the advantage of increased photon number, and thus better quantum noise reduction \cite{mccormick2008strong, glorieux2011quantum, liu2018quantum, thachil}. The bTMSS possess bipartite quantum correlations and can be efficiently generated with the existing technologies, making them good candidates for quantum-enhanced sensing \cite{ Dowran, lawrie, li2022quantum,  kamble, li2024harnessing}.    

The other major cornerstone in the field of quantum-enhanced metrology is quantum entanglement \cite{lloyd, tan, fink}. The nonlocal properties offered by quantum entanglement can produce advancement in sensing applications which is not possible with a single-mode squeezed state alone \cite{xiang2011entanglement, zhang2015entanglement, zhuang2017optimum, szigeti}. Various studies have already implemented the entanglement degree of freedom to improve the signal-to-noise ratio compared to the corresponding classical counterparts in different sensing tasks such as gyroscope \cite{fink}, target detection \cite{zhang2015entanglement}, phase estimation \cite{colangelo},  microscopy \cite{ono}, and many more \cite{huang2024entanglement}.

The conventional quantum sensing problems were dedicated in using the quantum principles to improve the precision of a single sensor. On the other hand, recently developed distributed quantum sensing (DQS) approach aims to improve the sensing of a global property like the weighted linear combination of multiple parameters of spatially distributed multiple sensors using entangled states \cite{proctor, Ge, Zhuang, xia1, Oh1, Guo, Xia2, zhuang_error, grace, Zhao, Zhang, Oh3, Liu, Sun, li2023,  Marino}. The DQS protocols for probing multiple spatially distributed sensors have been proposed and realized in both discrete-variable \cite{Ge, proctor} and continuous-variable  \cite{Zhuang, xia1, Oh1, Guo} regimes. It should be noted that DQS is different from multiparameter estimation in the sense that a multiparameter estimation problem aims at either estimating all the multiple unknown parameters or estimating a particular set of parameters, considering the rest as nuisance parameters \cite{Zhang}. However, in DQS, individual phases cannot be estimated with quantum-enhanced sensitivity; only their global property is recovered with utmost precision. The current development of quantum networks on a global scale, for example, satellite communications \cite{wang}, global clock synchronization\cite{komar}, and fiber optical communication networks \cite{sasaki} are some of the domains where the practical application of distributed sensing is relevant. 

\begin{figure}
    \centering
    \includegraphics[scale=0.85]{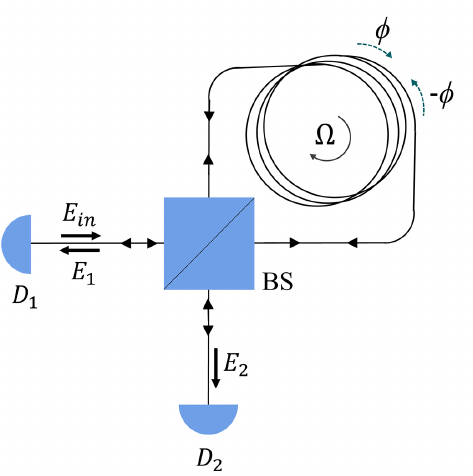}
    \caption{Optical gyroscope based on Sagnac interferometer with a 50:50 beam splitter (BS), rotating with an angular velocity of $\Omega$.}
    \label{fig:fig1}
\end{figure}

In the domain of quantum optical sensing, interferometry is one of the primary tools where the advantages of quantum-enhanced phase estimation are leveraged for various sensing applications \cite{caves, Yurke2, pooser1, Demkowicz, grangier, Holland, Weiping, marino2012effect}. One of the prominent applications of interferometry is optical gyroscopes based on the Sagnac effect that offers inertial navigation sensing for spacecraft, missiles, and satellites \cite{Sagnac, Vali, Chow, kolkiran, Jing, dowling1998, zhao2023quantum, Mehmet, xiao2020enhanced, wu2020atom}. In Fig.~\ref{fig:fig1}, we show a schematic of a classical optical gyroscope. A classical laser field $E_{in}$ is incident upon the beamsplitter (BS), which divides the field into two beams that travel in counterpropagating directions around the ring cavity. Due to the rotation of the gyroscope, they gain phase shifts $\phi$ and $-\phi$, and this phase information can be traced out from their interference. This phase shift of the two beams ($\Delta \phi=2\phi$) is directly related to the angular velocity $(\Omega)$ of the gyroscope as,
\begin{align}
    \Omega=\frac{\lambda c}{8 \pi A}\Delta\phi,
\end{align}
where A, $\lambda$, and $c$ are the area of the Sagnac loop, the wavelength of the laser used, and the speed of light inside the interferometer, respectively. Therefore, to improve the precision in estimating the angular velocity, one should seek methods to improve the phase sensitivity. However, with coherent probe states, the phase sensitivity is limited by the shot noise \cite{Jing}. Therefore, to improve the sensitivity further, one can make use of quantum resources \cite{Mehmet, Jing, xiao2020enhanced, wu2020atom, zhao2023quantum}. 

 Conventional inertial navigation systems use three orthogonal gyroscopes, with each gyroscope acting as an independent sensor to track the 3-dimensional motion. Distributed sensing enhances precision by deploying multiple gyroscopes per axis, enabling accurate phase measurements and improved rotation sensitivity.

In this work, we propose a scheme for quantum-enhanced phase sensing in a network of optical gyroscopes utilising bright two-mode squeezed states (bTMSS). In particular, we choose to work with bTMSS generated with a seeded input configuration, because such a configuration routinely has been realized via four-wave mixing (FWM) in an atomic vapor to generate bright quantum correlated twin beams (or, bTMSS) \cite{mccormick2008strong, glorieux2011quantum, Dowran, li2022quantum, li2024harnessing}, producing $\sim 10$ dB of squeezing. The reduced quadrature and intensity difference fluctuations of such beams offer a substantial signal-to-noise ratio owing to its increased photon number and bipartite quantum correlations. Therefore, using the bTMSS, we analyse two different configurations, one with distributing $M$-mode entangled states, and the other with $M$ separable states. We estimate the phase sensitivity in both configurations and optimize the parameters to achieve the maximum phase sensitivity.

The paper is organized as follows. In section II, we introduce the proposed configurations of distributed phase sensing in a network of optical gyroscopes using bTMSS. The subsections A, B, and C describe the analysis of phase sensitivity obtained with $M$-mode entangled bright beams, separate $M$ bTMSS, and the sensitivity ratio, respectively. In section III, we calculate the quantum Cramér-Rao bounds for the phase sensitivities of both configurations. We conclude in section IV by summarizing all the results, and finally provide the detailed calculations in an appendix in section V. 
\begin{figure*}[htbp]
    \centering
    \includegraphics[scale=0.76]{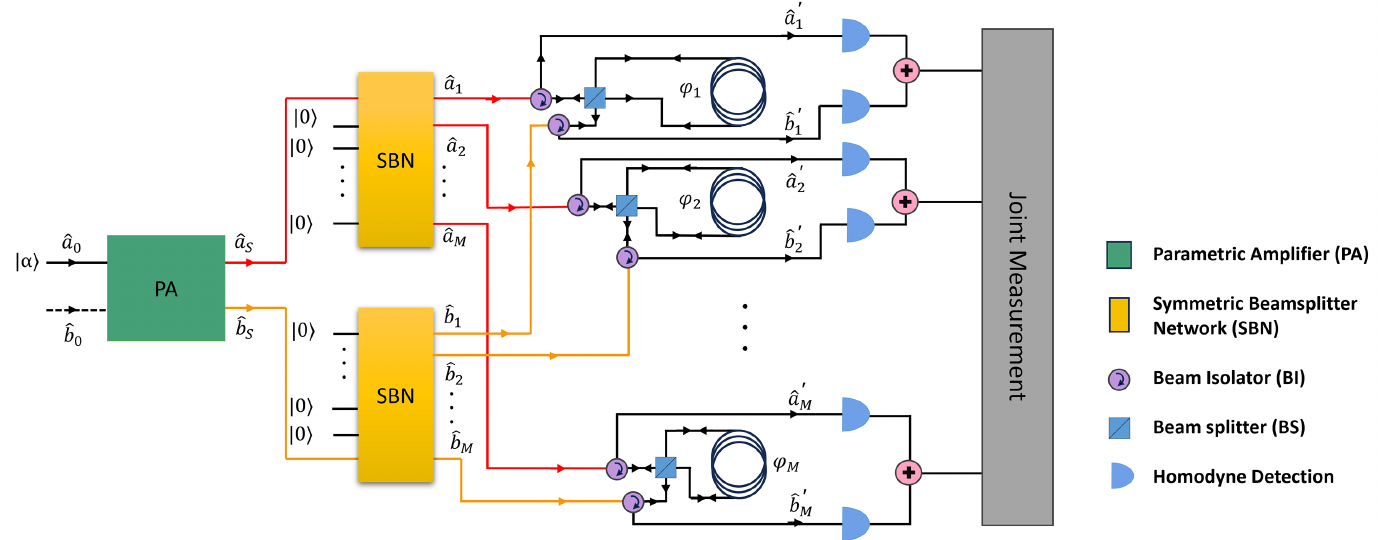}
    \caption{Distributed optical gyroscope configuration with mode-entangled probe states. The parametric amplifier with a squeezing parameter $r$ produces bright two-mode squeezed states in modes $\hat{a}_s$ and $\hat{b}_s$, which are given as inputs to two symmetric beamsplitter networks together with vacuum states in the remaining ports, producing M-mode entangled states at each SBN. They are further directed towards the optical gyroscopes via beam isolators. The beam isolators are beam routing optics (e.g., polarizing beamsplitters with appropriate waveplates) that isolate the input and output modes of the gyroscopes. Later, the phase-shifted output modes of the gyroscopes reach the detectors via the beam isolators, where a joint quadrature measurement is carried out.}
    \label{fig:fig2}
\end{figure*}
\section{Distributed phase sensing using \lowercase{b}TMSS IN optical gyroscopes}
Consider $M$ identical spatially separated optical gyroscopes, each measuring an unknown phase shift, as schematically shown in Fig.~\ref{fig:fig2}. The global parameter that we are interested in estimating is the average phase shift measured by the system as,
\begin{align}
    \phi=\frac{1}{M}\sum_{j=1}^M \phi_j.
\end{align}
Here, $\phi_j$ is the phase shift in the $j^{th}$ gyroscope. To estimate the average phase shift, we distribute the input states to probe the $M$ gyroscopes simultaneously and perform individual quadrature measurements at each output. 
In what follows, we analyse the phase sensitivity in two configurations: the one in which $M$ optical gyroscopes are probed with an $M$-mode entangled bright beams, and the other in which the same set of sensors is probed with independent $M$ bright TMSSs.

\subsection{With M-Mode Entangled bTMSS}
In this scheme, optical gyroscopes are probed by continuous variable mode-entangled states. The configuration is illustrated in Fig.~\ref{fig:fig2}. We employ a parametric amplifier (PA) with squeezing parameter $r$, which produces two-mode squeezed states in modes $\hat{a}_s$ and $\hat{b}_s$. We seed the input mode $\hat{a}_0$ of the parametric amplifier with a coherent state of amplitude $\alpha$, and keep the mode $\hat{b}_0$ vacuum, generating bright twin beams in the modes $\hat{a}_s$ and $\hat{b}_s$ as,
 \begin{align}
     \hat{a}_s=\hat{a}_0\cosh{r} + \hat{b}_0^\dag\sinh{r}, \nonumber \\
     \hat{b}_s=\hat{b}_0\cosh{r}+\hat{a}_0^\dag\sinh{r},
 \end{align}
 where $\cosh{r}$ is the gain of the nonlinear process in the PA. These bTMSS are further sent to two lossless $M \times M$ symmetric beamsplitter networks (SBN) having $M$ input and $M$ output ports, with their remaining $M-1$ input ports at vacuum. The SBNs are designed such that each input mode $\hat{a}_s$ ($\hat{b}_s$) of the SBN equally contributes to the $M$ output modes $\hat{a}_j$s ($\hat{b}_j$s) such that,  
\begin{align}
    \hat{a}_s=\frac{1}{\sqrt{M}} \sum_{j=1}^M \hat{a}_j,  ~~~~~~~
    \hat{b}_s=\frac{1}{\sqrt{M}} \sum_{j=1}^M \hat{b}_j.
\end{align} 
Consequently, the two initial modes $\hat{a}_s$ and $\hat{b}_s$, which had strong bipartite quantum correlations among them, are mixed with vacuum modes at their respective SBNs to project them into $M$ output modes $\hat{a}_j$s and $\hat{b}_j$s that possess continuous variable multipartite entanglement across their $M$ modes. These $M$-mode entangled states $\hat{a}_j$s and $\hat{b}_j$s are further passed through the inputs of the optical gyroscopes via the transmission port of the beam isolators (BI). The BIs are passive beam routing optics like the polarization-based devices that isolate the input and output beams of the gyroscopes. The beams in the modes $\hat{a}_j$ and $\hat{b}_j$ travel in a counterpropagating direction inside the $j^{th}$ gyroscope and due to the rotation of the gyroscopes, they gain the phase shifts $\phi_j$ and $-\phi_j$, respectively. The resultant phase-shifted modes become,
 \begin{align}
     \Tilde{\hat{a}}_j=\cos{\phi_j} ~\hat{a}_j - i~ \sin{\phi_j} ~\hat{b}_j, \nonumber \\
     \Tilde{\hat{b}}_j=\cos{\phi_j} ~\hat{b}_j - i~ \sin{\phi_j} ~\hat{a}_j.
 \end{align}
Moreover, for simplifying the calculations, we assume that the unknown phase shifts are small enough such that $\sin{\phi} \approx \phi$. This approximation to the linear regime holds well in the practical applications scenes of inertial navigation and precision metrology tasks, where small phase shifts are measured and integrated over time to obtain the total rotation. 

 In addition, we also consider internal photon loss by assuming lossy channels of transmissivity $\eta$, which leads to the transformation,
\begin{align}
    \hat{a}_j^\prime=\sqrt{\eta}~\Tilde{\hat{a}}_j + \sqrt{1-\eta}~\hat{a}_{j,vac}, \nonumber \\
    \hat{b}_j^\prime=\sqrt{\eta}~\Tilde{\hat{b}}_j + \sqrt{1-\eta}~\hat{b}_{j,vac}, 
\end{align} 
 where $\hat{a}_{j, vac}$ and $\hat{b}_{j, vac}$ are the arbitrary vacuum modes coupling through the unused ports of the fictitious beamsplitters with transmissivity $\eta$ considered to model the lossy channel. 
 
 Further, the phase-shifted modes, $\hat{a}_j^\prime$s and $\hat{b}_j^\prime$s, return via the reflection port of the BIs and reach the output detectors. We then measure the phase quadratures of each output mode independently through homodyne detection, and collectively combine them for a joint quadrature measurement.  We define the superposition quadrature of the output modes of individual optical gyroscopes as, 
 \begin{align}
    Y_j=\frac{Y_{a_j^{\prime}}+Y_{b_j^{\prime}}}{\sqrt{2}},
\end{align}
 which allows for leveraging the advantage of the correlations between the modes of the two-mode squeezed state.
 Finally, we use the following joint quadrature as the estimator for the sensing task, 
\begin{align}
    Y_+=\sum_{j=1}^M Y_j .
\end{align}
 
Using Eqs.~(2-8), we calculate the expectation value and variance of the joint phase quadrature as (detailed calculations are given in Appendix A),
\begin{gather}
    \braket{Y_+}=-\sqrt{\eta M}\alpha e^r \phi, \\
    \Delta Y_+^2=\frac{M}{2}(1+\eta e^{-2r}-\eta).
\end{gather}
In order to estimate the average phase shift, we make use of the linear error propagation formula, and obtain the uncertainty in the average phase measurements as,
\begin{align}
    \Delta\phi^2=\frac{\Delta Y_+^2}{\big|{\partial \braket{\hat{Y}_+}}/{\partial \phi}\big|^2}.
\end{align}
In the analysis of the subsequent sections, we define the variance $\Delta\phi^2$ as the phase sensitivity; thus, the smaller the value of $\Delta\phi^2$, the better the phase sensitivity. 

Now, using Eq.~(9) and Eq.~(10) in Eq.~(11), we calculate the phase sensitivity corresponding to the mode-entangled configuration shown in Fig.~\ref{fig:fig2} as, 
\begin{align}
    \Delta\phi^2_{ent}= \frac{e^{-2r}+1/\eta -1}{2\alpha^2 e^{2r}}.
\end{align}
From Eq.~(12), it is evident that phase sensitivity is independent of the number of sensor nodes. It is equivalent to the phase sensitivity of a single sensor with the same resources.
 
The phase sensitivity given in Eq.~(12) depends on various factors, including the amplitude of the input coherent state $(\alpha)$, the squeezing parameter $(r)$ of the parametric amplifier, and the transmissivity $(\eta)$. Therefore, for a given $\eta$, the phase sensitivity can be optimized by controlling $\alpha$ and $r$; the factors that are eventually deciding the number of photons probing each gyroscope. For that, we follow the Lagrangian multiplier formalism with the constraint that the average number of photons probing each gyroscope is $N$. We calculate the total number of photons sensing the whole system as,
\begin{align}
     N_{tot}=\alpha^2 \cosh{2r}+2\sinh^2{r}.
\end{align}
For optimizing the sensitivity by controlling $r$ and $\alpha$, with a constraint on the average photon number per gyroscope, we define the Lagrange function as,
\begin{align}
    \mathcal{L}(\alpha,r,\lambda) = \Delta\phi^2_{ent} + \lambda(\frac{\eta N_{tot}}{M}- N).
\end{align}
On solving Eq.~(14) (see Appendix A for the details), we obtain an implicit expression connecting $r$ with $N$,
\begin{gather}
    \eta  e^{2 r} \left( 2(M N-1)(2 e^{2 r} -e^{4 r})-6 M N+e^{8 r}+1\right)  \nonumber \\
    -4 M N e^{4 r}-\left(e^{2 r}-1\right)^2
   \left(3 e^{2 r}+e^{6 r}-2\right) \eta^2=0.
\end{gather}
Also, for $\alpha$, we obtain the expression,
\begin{gather}
  \alpha=  \sqrt{\frac{ 2e^{2 r} (M N-\eta \cosh{2r}+\eta)}{\eta\left(e^{4 r}+1\right) }}.
\end{gather}
Using Eqs.~(15) and (16), we can calculate the optimal values of $r$ and $\alpha$ that minimize the variance in Eq.~(12) for given values of $M, N$ and $\eta$. Since Eq.~(15) is an implicit equation, we calculate the optimal value of $r$ numerically. Figure \ref{fig:fig2} shows the optimal values of (a) $r$ and (b) $\alpha$, respectively, as a function of both $M$ and $N$, for the case when the photon transmissivity, $\eta=0.9$.  
\begin{figure}
    \centering
    \includegraphics{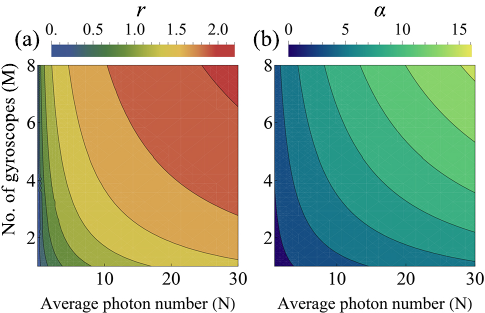}
    \caption{Optimal values of (a) squeezing parameter, $r$, and (b) coherent seed amplitude, $\alpha$, as a function of average photon number, $N$ and number of gyroscope units, $M$, at a given value of transmissivity, $\eta=0.9$.}
    \label{fig:fig3}
\end{figure}

It is evident from the Fig.~\ref{fig:fig2} that the optimal values of $r$ and $\alpha$ required to optimize the phase sensitivity increase with increasing $M$ and $N$. Also, it can be seen that with $10\%$ photon loss in every channel in the system, the optimal values of $r$ and $\alpha$ are in the practically achievable ranges using the current technologies. For example, given $M=4, ~N=10,$ and $\eta=0.9$, then the optimal value of $r$ and $\alpha$ are $1.53$ and $1.81$, respectively. 
\begin{figure}
    \centering
    \includegraphics{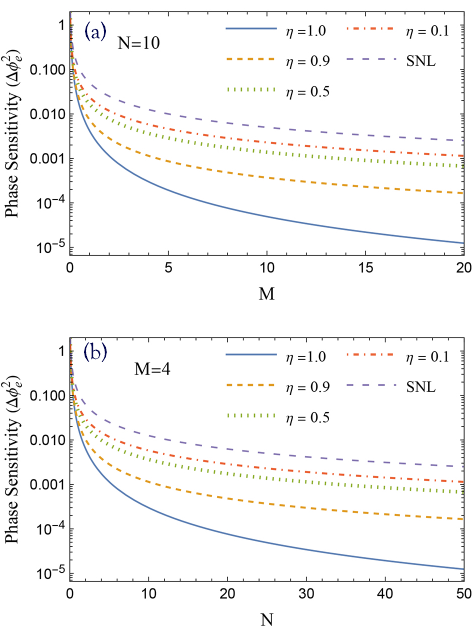}
    \caption{Optimized Phase sensitivity $\Delta\phi_e^2$ of the mode-entangled configuration as a function of (a) the number of sensors $M$, at a constant value of average photon number per sensor $N=10$; (b) average photon number per sensor $N$, at a fixed number of sensors $M=4$. In both (a) and (b), the purple dashed trace represents the SNL of the phase sensitivity. The blue solid trace, the orange dashed trace, the green dotted trace, and the red dotted-dash trace represent the phase sensitivity corresponding to $\eta=1.0$, $\eta=0.9$, $\eta=0.5$, and $\eta=0.1$, respectively.}
    \label{fig:fig4}
\end{figure}

The optimal values of $r$ and $\alpha$ are then substituted back in Eq.~(12) to obtain the optimized phase sensitivity represented by $\Delta\phi^2_e$, and it is plotted in Fig.~\ref{fig:fig4}. Figures~\ref{fig:fig4} (a) and (b) show $\Delta\phi^2_e$ as a function of the number of gyroscopes ($M$) and the average number of photons ($N$) probing each gyroscope, respectively. We keep $N=10$ in  Fig.~\ref{fig:fig4} (a) and $M=4$ in Fig.~\ref{fig:fig4} (b), and analyze the results for different $\eta$ values of $1.0$, $0.9$, $0.5$ and $0.1$. To present the quantum enhancement in phase sensitivity obtained by the system, we compare $\Delta\phi^2_e$ with the shot-noise limit (SNL) in sensitivity obtained when only coherent states probe the system. To obtain the SNL in the phase sensitivity, we put $r=0$ in Eq.~(12) and also use the constraint on average photon number, which provides the SNL as $1/2M N$, and is plotted in Fig.~\ref{fig:fig4} (purple dashed trace). In both the plots in Figs.~\ref{fig:fig4} (a) and (b), the blue solid trace represents the phase sensitivity from the mode-entangled configuration corresponding to $\eta=1.0$, the orange dashed trace represents $\eta=0.9$, the green dotted trace represents $\eta=0.5$, and the red dotted-dash trace represents $\eta=0.1$. We can see from the plots in both Figs.~\ref{fig:fig4} (a) and (b) that the phase sensitivity improves with increasing $M$ and $N$. Also, at any transmissivity, the sensitivity of the mode-entangled configuration surpasses the SNL. 
\begin{figure}
    \centering
    \includegraphics[scale=0.7]{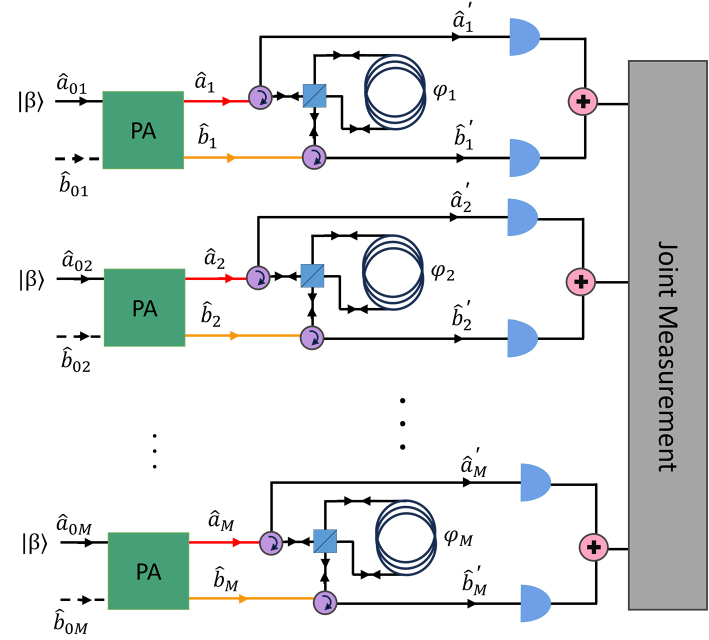}
    \caption{Distributed optical gyroscope configuration with $M$ separate bTMSS as input states. The M parametric amplifiers, each with the same squeezing parameter $r$, produce bright two-mode squeezed states in modes $a_j$ and $b_j$ ($1 \le j \le M$), which are given as inputs to separate M optical gyroscopes. The phase-shifted modes reach the detectors via the beam isolators, where a joint quadrature measurement is carried out.}
    \label{fig:fig5}
\end{figure}

\subsection{With M-Mode Separate bTMSS}
\begin{figure}
    \centering
    \includegraphics{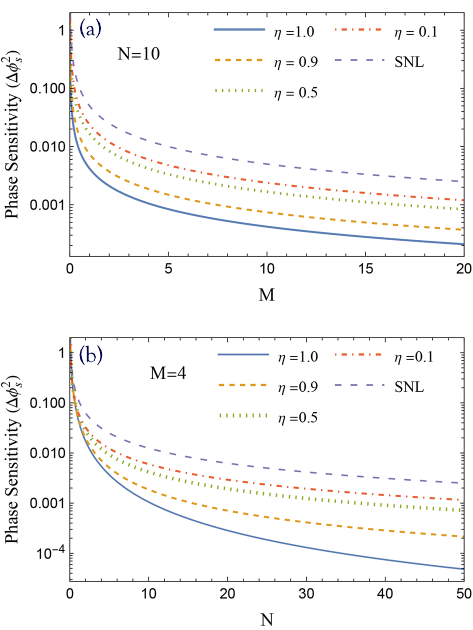}
    \caption{Optimized Phase sensitivity $\Delta\phi_s^2$ of the separable configuration as a function of (a) the number of sensors $M$, at a constant value of average photon number per sensor $N=10$; (b) average photon number per sensor $N$, at a fixed number of sensors $M=4$.  In both (a) and (b), the purple dashed trace represents the SNL of the phase sensitivity. The blue solid trace, the orange dashed trace, the green dotted trace, and the red dotted-dash trace represent the phase sensitivity corresponding to $\eta=1.0$, $\eta=0.9$, $\eta=0.5$, and $\eta=0.1$, respectively.}
    \label{fig:fig6}
\end{figure}

In order to compare the enhancement obtained from the M-mode entangled probe state with that from an $M$-mode unentangled probe state, we study the system as probed by $M$ independent bTMSS. For that, we employ $M$ separate PAs, each with a coherent seed of amplitude $\beta$, generating bTMSS as input to separate optical gyroscopes, as shown in Fig.~\ref{fig:fig5}. With this configuration, the phase sensitivity obtained is (for detailed calculation, see Appendix B),
\begin{align}
    \Delta\phi^2_{sep}=\frac{e^{-2r}+1/\eta -1}{2\beta^2 M e^{2r}}.
\end{align}
From Eq.~(17), it can be seen that the phase sensitivity improves with increasing number of gyroscopes. For a fair comparison of the sensitivity with the mode-entangled case, we consider the same average number of photons probing each gyroscope as $N$.  Again, using Lagrangian multiplier formalism, we obtain the optimal values of $\beta$ and $r$ that minimize the phase sensitivity, bound to the average photon number constraint (detailed calculation is given in Appendix B). Substituting the optimal values of $\beta$ and $r$ into Eq.~(17), we obtain the optimized phase sensitivity $\Delta\phi^2_s$ for the separable configuration, and plot in Fig.~\ref{fig:fig6}. With $r=0$ in Eq.~(17) and using the average photon number constraint, we obtain the SNL in sensitivity of this configuration as $1/2M N$, same as the SNL in the mode entangled configuration.

We plot the optimized phase sensitivity $\Delta\phi^2_s$ for the separable configuration along with the SNL as a function of $M$ and $N$ in Figs.~\ref{fig:fig6} (a) and (b), respectively. We keep $N=10$ in Fig.~\ref{fig:fig6} (a), and $M=4$ in Fig.~\ref{fig:fig6} (b), and analyze the sensitivity for different values of transmissivities. In both plots, the purple dashed trace represents the SNL, the solid blue represents optimized phase sensitivity when $\eta=1.0$, the orange dashed represents $\eta=0.9$, the green dotted, $\eta=0.5$, and the red dotted-dash, $\eta=0.1$. Here also, we can observe that the separable configuration can always beat the SNL at any value of the transmissivity. 
\begin{figure}
    \centering
    \includegraphics{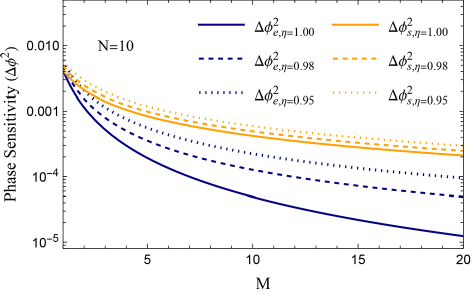}
    \caption{Optimized phase sensitivities of the mode-entangled $(\Delta\phi^2_e)$ and separable $(\Delta\phi^2_s)$ configurations  as a function of number of gyroscopes, $M$, with average photon number $N=10$. The solid traces correspond to the lossless condition, the dashed traces represent the lossy condition with $\eta=0.98$, and the dotted traces represent $\eta=0.95$. For all the loss conditions plotted in the graph, the blue traces correspond to the phase sensitivities of the mode-entangled configuration, and the orange to the separable configuration.}
    \label{fig:fig7}
\end{figure}

In order to compare the results of the two configurations shown in Fig.~\ref{fig:fig2} and Fig.~\ref{fig:fig5}, we plot the optimized phase sensitivities of both mode-entangled and separable configurations in Fig.~\ref{fig:fig7} (blue traces: mode-entangled configuration, orange traces: mode separable configuration) at different loss conditions, as a function of number of gyroscope units, with an average photon number per gyroscope, $N=10$. We can see from the plots that the mode-entangled scheme provides better sensitivity than that with the separable configuration. Moreover, even in the moderate lossy condition ($\eta=0.95$), the sensitivity for the mode-entangled configuration (blue dotted trace) can beat the sensitivity with the ideal no-loss condition of the separable configuration (orange solid trace). 

\subsection{Sensitivity Ratio}
To compare the enhancement in the sensitivity obtained by the use of mode-entangled states over the separable one, we define a sensitivity ratio ($R$) of both cases,
\begin{align}
    R=\frac{\Delta\phi^2_s}{\Delta\phi^2_e}.
\end{align}
Therefore, $R > 1$ signifies the advantage of using the configuration with M-mode entangled states over the same number of separable bTMSS.

\begin{figure}
    \centering
    \includegraphics{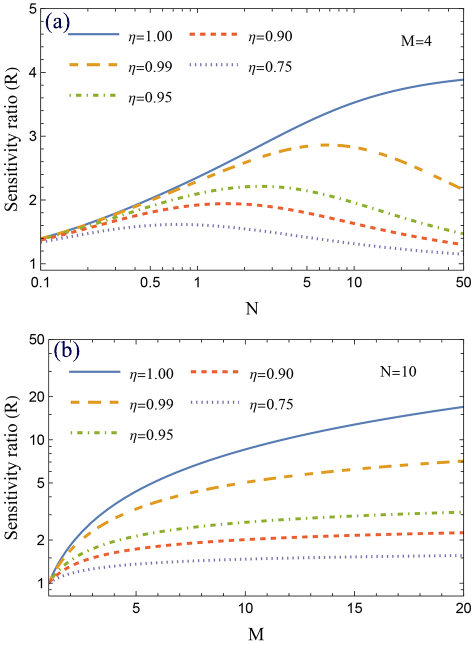}
    \caption{(a) Sensitivity ratio $(R)$ as a function of average photon number $N$, keeping the number of gyroscopes $M=4$, plotted at different loss conditions. In every case, the sensitivity ratio $R$ is greater than unity, indicating enhanced performance of the mode-entangled configuration over the separable case. In the no-loss case, the ratio $R$ increases with increasing $N$. For non-zero losses, the performance gain peaks at particular values of $N$ and further degrades to unity, showing comparable performance between both configurations. (b) Sensitivity ratio as a function of number of gyroscopes, $M$, keeping the average photon number, $N=10$, at different loss conditions. For every case, the performance of the mode-entangled configuration outperforms the separable configuration as $M$ increases; however, the rate is lower with increasing losses.}
    \label{fig:fig8}
\end{figure}

We plot the sensitivity ratio, $R$, as a function of the number of photons probing each gyroscope, $N$, in Fig.~\ref{fig:fig8} (a), by keeping $M=4$. For a given value of photon number $(N)$ per gyroscope, in both the configurations, the optimization process finds the optimal values of the squeezing parameter ($r$) and the coherent seed amplitude ($\alpha$ for mode-entangled configuration and $\beta$ for mode-separable configuration) that minimizes the uncertainty in the average phase. We can see from Fig.~\ref{fig:fig8} (a) that when there is no internal loss, the entangled approach has a definite enhancement over the separable approach with increasing values of $N$. However, for the lossy cases, the performance of the entangled approach peaks at a particular value of $N$ and is seen to be slowly degrading and tends to unity, that is, it follows the same trend as the separable approach for large photon numbers. This is due to the fact that in the presence of internal photon loss, the correlations in the mode-entangled state are reduced, making it deviate from its enhanced performance. 

As the losses increase, the finite value of $N$, at which the sensitivity ratio peaks shifts towards the lesser $N$ value. This indicates that when there are more photon losses, the mode-entangled configuration loses its supremacy over the separable case at lesser values of $N$. For example, when the channel transmissivity is $99\%$ in every photon channel, the sensitivity ratio peaks at $N=6.7$, with an enhancement factor of of $2.86 ~(4.56$ dB) over the separable configuration. Similarly, when the transmissivity is $95\%$, the ratio peaks at $N=2.53$ with peak value of enhancement factor $2.21$ ~($3.45$ dB). Therefore, it is optimal to operate the configuration at these values of $N$, where one can exploit the most advantage of the mode-entangled configuration over the separable one.
Also, one can calculate the quantum enhancement in the phase sensitivity of the mode-entangled configuration compared to the corresponding SNL. For example, when $\eta=0.99$ and $N=6.7$, $\Delta\phi^2_{SNL}/\Delta\phi^2_e$ is $23.22$ giving a $13.66$ dB quantum enhancement in phase sensitivity. Similarly, when $\eta=0.95$, the sensitivity ratio peaks at $N=2.53$, which corresponds to an initial squeezing of $9.8$ dB, the quantum enhancement obtained is $9.28$ dB.

Furthermore, we plot the behavior of the sensitivity ratio with respect to the number of gyroscopes $(M)$ in Fig.~\ref{fig:fig8} (b), with a constant $N=10$. It can be seen that the sensitivity ratio always increases with an increase in the number of gyroscopes. However, the rate of increase reduces with internal losses. From Fig.~\ref{fig:fig8} (a) and (b), one can see that as the photon loss increases, the performance of both the mode-entangled and the separable configurations becomes comparable. However, there is still a practical advantage for the entangled approach that only a single parametric amplifier is required to produce the same sensitivity as produced by the $M$ parametric amplifiers in the separable approach.

\section{QUANTUM CRAMÉR-RAO BOUND}
For every sensing problem, the ultimate precision in estimating an unknown parameter is upper bounded by the Quantum Cramér-Rao Bound (QCRB), calculated using the Quantum Fisher Information (QFI) obtained by maximizing over all possible measurements. For the case of estimation of multiple parameters $\{\phi_1, \phi_2,...\phi_M\}$, the quantum-enhanced estimation error for an unbiased estimator $\boldsymbol{\phi}$ is lower bounded by the multiparameter Quantum Cramér-Rao Bound, defined by the inverse of the quantum Fisher information matrix (QFIM) \cite{helstrom}. Accordingly, the uncertainty in the estimation of a linear combination of multiple parameters,  $\boldsymbol{\phi}=\sum_{i=1}^M w_i \phi_i$, is bounded as,
\begin{align}
    \Delta {\phi}^2 \ge w^T F^{-1} w=\Delta {\phi}^2_{cr},
\end{align}
where $F$ is the QFIM of the system and $w=\{w_1, w_2,...w_M\}$ is the weight coefficients. Therefore, any sensing task can be further improved up to its ultimate precision bound given by the QCRB using optimal probe states and measurement schemes. In our case of estimation of the average of $M$ unknown phase shifts, the weight coefficients are all equal to  $w_i=M$ for $1\le i\le M$. In the following, we are determining the quantum Fisher information matrix of our distributed gyroscope configurations using both $M$-mode entangled states, as well as separate $M$ bTMSS states. 

\subsection{QCRB for the Mode-Entangled Configuration}

The mode-entangled configuration shown in Fig.~\ref{fig:fig2} is a continuous-variable Bosonic Gaussian system. The input probe state $\rho$ can be fully characterized by the displacement vector $\textbf{d}$ and the covariance matrix $V$, defined in terms of the quadrature vector $\hat{\textbf{r}} =\{\hat{x}_1,\hat{y}_1,...\hat{x}_M,\hat{y}_M\}^T$, as,
\begin{gather}
    \textbf{d}=\braket{\hat{\textbf{r}}},   \nonumber   \\
    V_{jk}=\frac{1}{2}\braket{\Delta \hat{r}_j \Delta \hat{r}_k+ \Delta \hat{r}_k \Delta \hat{r}_j}.
\end{gather}
Here, $\Delta \hat{r}_j=\hat{r}_j-\braket{\hat{r}_j}$ and $[\hat{r}_j,\hat{r}_k]=i\Omega_{jk}$, with $\Omega$ given by,
\begin{gather}
    \Omega= \bigoplus_{k=1}^{M}  \begin{pmatrix} 0 & 1 \\ -1 & 0 \end{pmatrix}=\bigoplus_{k=1}^{M}\omega. 
\end{gather}
After the multiphase encoding, the input state $\hat{\rho}$ transforms into $\hat{\rho}_{\phi}$ with displacement vector, $\boldsymbol{d}_{\phi}$ and covariance matrix, $V_{\phi}$ . The displacement vector, $\boldsymbol{d}_{\phi}$ of dimension $2M$ can be written as a block vector of $M$ subvectors of dimension $2\times1$, with the $i^{th}$ $(1\le i\le M)$ block given by,
\begin{align}
    \boldsymbol{d}_\phi^{i}= \sqrt{\frac{\eta}{M}}e^{r}\alpha \begin{bmatrix}
      \cos{\phi_i} \\ -\sin{\phi_i} 
    \end{bmatrix}.
\end{align}
Similarly, the symmetric covariance matrix $V_\phi$ can be written as an $M\times M$ block matrix, with diagonal and off-diagonal $2\times2$ submatrices taking the form, 
\begin{align}
    V_\phi^{(i,i)}=\frac{1}{2}\begin{bmatrix}
    p+q \cos{2\phi_i}+1 & q \sin{2\phi_i}\\
    q \sin{2\phi_i}&p-q \cos{2\phi_i}+1\\   
\end{bmatrix}, \nonumber
\end{align}
\begin{align}
    V_\phi^{(i,j)}=\frac{1}{2}\Large\begin{bmatrix}
    \substack{p \cos{(\phi_i-\phi_j)}+\\q \cos{(\phi_i+\phi_j)}} &  &\substack{p \sin{(\phi_j-\phi_i)}+\\q \sin{(\phi_i+\phi_j)}}\\ & &\\
   \substack{p \sin{(\phi_i-\phi_j)}+\\q \sin{(\phi_i+\phi_j)}} &  &\substack{p \cos{(\phi_i-\phi_j)}\\-q \cos{(\phi_i+\phi_j)}}\\
\end{bmatrix}.
\end{align}
Here, $p=2\eta \sinh^2{r}/M$ and $q=\eta \sinh{2r}/M$. 

 If we consider no photon losses in the system, then $\rho_{\phi}$ is an $M$-mode pure Gaussian state with isothermal photon number equal to zero. For an $M$-mode pure Gaussian state, we may write the QFI matrix as \cite{banchi},
\begin{align}
     F_{ij}=Tr\big[\Omega\frac{\partial V_\phi}{\partial\phi_i}\Omega\frac{\partial V_\phi}{\partial\phi_j}\big]+\frac{\partial \boldsymbol{d}_\phi^T}{\partial\phi_i}V_\phi^{-1}\frac{\partial \boldsymbol{d}_\phi}{\partial\phi_j}.
\end{align}
The QFIM is independent of $\boldsymbol{\phi}$ under unitary transformation. Therefore, we can set $\boldsymbol{\phi}=0$ without loss of generality, and Eq.~(24) can be simplified as in \cite{Oh1},
\begin{align}
    F_{ij}=2 Tr \big[V^{(i,j)} V^{(j,i)}\big]-\delta_{ij}+(\omega \boldsymbol{d^{i}})^T\big[V^{-1}\big]^{(i,j)}\omega \boldsymbol{d^{j}}.
\end{align}
Consequently, using Eq.~(22) and Eq.~(23) in Eq.~(25), we obtain the elements of the QFIM as,
\begin{gather}
    F_{ii}=F_{11}=\frac{1}{M^2} \big[4\sinh^2{r}\cosh{2r}+4M\sinh^2{r}~ +  \nonumber \\
    2\alpha^2e^{4r}\{(M-1)(e^{-2r}-1)+M\}\big],~~~~ \forall ~i ,\nonumber \\
    F_{ij}=F_{12}=\frac{1}{M^2}\big[4\sinh^2{r}\cosh{2r}-2\alpha^2e^{4r}(e^{-2r}-1)\big], \nonumber \\
    \forall ~i \ne j.
\end{gather}     
Now, using Eq.~(26) in Eq.~(19), we obtain the QCRB for the mode-entangled configuration in the no-loss case as,
\begin{gather}
    \Delta\phi^2_{ent, \bar{cr}}=\frac{1}{M}\big[F^{-1}_{11}+(M-1)F^{-1}_{12}\big] \nonumber \\
    =\frac{1}{M}\frac{1}{\big[F_{11}+(M-1)F_{12}\big]} \nonumber \\
    =\frac{1}{2[e^{4r}\alpha^2+\sinh^2{2r}]}.
\end{gather}
It can be seen from Eq.~(27) that the QCRB for the mode-entangled configuration with $M$ gyroscopes is independent of $M$. 

Furthermore, to obtain the results corresponding to a general lossy case, we cannot use the formula given in Eq.~(24); because, in the presence of losses, the state $\rho_{\phi}$ is no longer an isothermal state. Therefore, we use the procedures given in \cite{Nichols, pirandola2009correlation} to formulate the QFI matrix numerically and use Eq.~(19) to calculate the QCRB in phase sensitivity (see Appendix C for the details). Due to the computational complexity in the methods given in \cite{Nichols}, we first formulate the QCRBs corresponding to the simpler cases of $M=1$ and $M=2$. The presence of vacuum inputs in the $(M-1)$ ports of the beam splitter network yields singular results in the formulation, as pointed out in \cite{Safranek, Guo}. To resolve this numerical problem, we adopt the same pragmatic technique used in \cite{Guo} of introducing very weak thermal states of mean photon number of order $10^{-6}$ at the vacuum input ports of the SBNs. These factors can be easily ignored in the final expression of the QCRB, which takes the following form for both cases of $M=1$ and $M=2$ of the mode-entangled configuration as, 
\begin{align}
    \Delta\phi^2_{ent, cr}=\frac{(1-\eta+\eta e^{2r})(1-\eta+\eta e^{-2r})}{2\eta[e^{2r}\alpha^2(1-\eta+\eta e^{2r})+\eta \sinh^2{2r}]}.
\end{align}
Since for both cases of $M$, the QCRB obtained has the same expression, it implies that the QCRB expression of the mode-entangled configuration as obtained in Eq.~(28) is general for any value of $M$. Furthermore, when there are no losses in the system, the QCRB becomes
\begin{align}
    \Delta\phi^2_{ent, cr|\eta \to 1}=\frac{1}{2[e^{4r}\alpha^2+\sinh^2{2r}]},
\end{align}
which is the same as the result obtained in Eq.~(27). Also, for the limit ($e^{4r}\alpha^2 \gg  \sinh^2{2r}$), Eq.~(29) is equivalent to the result in Eq.~(12) when $\eta \to 1$. This implies that if the coherent seed amplitudes are large enough, the mode-entangled configuration can saturate the QCRB and obtain ultimate precision in phase sensitivity. 

\begin{figure}
    \centering
    \includegraphics{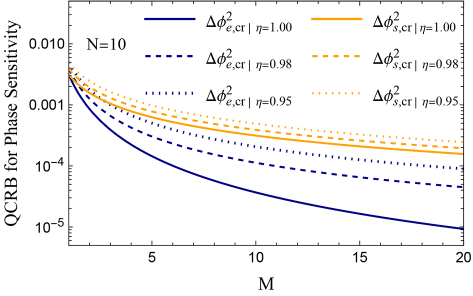}
    \caption{Optimized Quantum Cramér-Rao Bound of phase sensitivities obtained from the mode-entangled and separable configurations, plotted as a function of the number of gyroscope units, $M$, with average photon number, $N=10$, at different loss conditions. The solid traces correspond to the lossless condition $\eta=1.00$, the dashed traces represent the lossy condition with $\eta=0.98$, and the dotted traces represent $\eta=0.95$. For all the loss conditions plotted in the graph, the blue traces correspond to the phase sensitivities of the mode-entangled configuration, and the orange to the separable configuration. }
    \label{fig:fig9}
\end{figure}

Now, as in the previous formulations done in parts A and B of section (II), the QCRB can also be optimized for $\alpha$ and $r$ using the Lagrangian multiplier formulation, obeying the average photon number constraint. This leads to the final optimized QCRB of the phase sensitivity of the mode-entangled configuration, $\Delta\phi^2_{e, cr}$, which we calculate numerically, and is plotted in Fig.~\ref{fig:fig9} as a function of the number of gyroscopes, $M$, with blue traces. We can see from the figure that the optimized QCRB has an inverse scaling with $M$.   
\begin{figure}
    \centering
    \includegraphics{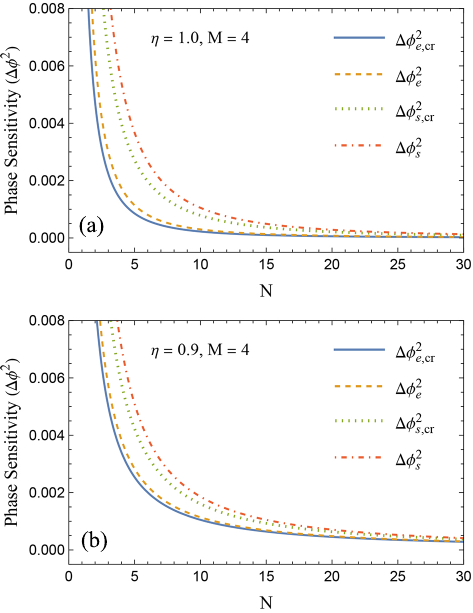}
    \caption{Optimized Sensitivities as a function of average photon number, $N$, when (a) $\eta=1.0$ and (b) $\eta=0.9$, provided the number of gyroscopes is $M=4$. In both plots, the blue solid trace and the orange dashed trace correspond to the optimized QCRB and the optimized sensitivity obtained using the homodyne measurement described in section II, respectively, of the mode-entangled configuration. Similarly, the green dotted trace and the red dash-dotted trace represent the optimized QCRB and the optimized sensitivity corresponding to the homodyne measurement scheme, respectively, for the separable configuration.}
    \label{fig:fig10}
\end{figure} 
\subsection{QCRB for the Separate bTMSS Configuration}
Similarly, we calculate the quantum Cramér-Rao bound for the phase sensitivity for the separable distributed gyroscope configuration shown in Fig.~\ref{fig:fig5} with $M$ gyroscope units and coherent input of amplitude $\beta$. We obtain the QCRB of this configuration as,
\begin{align}
    \Delta\phi^2_{sep, cr}=\frac{(1-\eta+\eta e^{2r})(1-\eta+\eta e^{-2r})}{2\eta M[e^{2r}\beta^2(1-\eta+\eta e^{2r})+\eta \sinh^2{2r}]}.
\end{align}
When, $\eta \to 1$, the phase sensitivity become,
\begin{gather}
    \Delta\phi^2_{{sep, cr}{|_{\eta \to 1}}}=\frac{1}{2M[e^{4r}\beta^2+ \sinh^2{2r}]}, 
\end{gather}
equivalent to the result in Eq.~(17), provided $e^{2r}\beta^2\gg\sinh^2{2r}$. 

Again, optimizing the sensitivity for $\beta$ and $r$, constraining the average number of photons in each gyroscope to be $N$, we numerically obtain the final optimized sensitivity of the separable configuration as $\Delta\phi^2_{s,cr}$.  

In Fig.~\ref{fig:fig9}, we plot the optimized QCRBs of the mode-entangled $\Delta\phi^2_{e, cr}$ as well as the separable configuration $\Delta\phi^2_{s, cr}$ at three different loss conditions as a function of the number of gyroscope units, $M$, keeping the average photon number per gyroscope, $N=10$. It is evident from the plots that the mode-entangled configuration outperforms the separable configuration at every loss condition.  

We also compare the optimized QCRBs with the optimized sensitivity results that are obtained using the measurement technique described in Section II. We plot the four optimized sensitivity results in Fig.~\ref{fig:fig10} for $M=4$, at two different loss conditions (a) $\eta=1.0$ and (b) $\eta=0.9$. 
In both cases, it follows that $\Delta\phi^2_{e, cr} <   \Delta\phi_e^2 < \Delta\phi^2_{s, cr}  < \Delta\phi_s^2$. It shows that, using the proposed mode-entangled configuration, one can obtain average phase sensitivity for a spatially distributed $M$ gyroscopes better than the ultimate precision possible with a separable configuration.

Moreover, at larger values of $N$, the sensitivity with the proposed configurations is approaching the ultimate precision limits, in both the no-loss case as well as the moderate lossy case, as can be seen from Fig.~\ref{fig:fig10} (a) and (b), respectively. Such a higher value of $N$ can be easily achieved with bright TMSSs.
\section{Distributed phase sensing in optical gyroscopes using double-seeded Parametric Amplifiers}
In the proposed distributed configurations shown in Fig.~\ref{fig:fig2} and Fig.~\ref{fig:fig5}, bTMSS are generated using PAs that are single-seeded with coherent states. On the other hand, we can also think of a configuration in which both the input modes of the PA are seeded. Consider the PA in the mode-entangled configuration in Fig.~\ref{fig:fig2} is seeded by coherent states of equal amplitude $\alpha$ to both the input ports. Then, using the similar formulations used in section (II), we obtain the sensitivity for the average phase shift as,
\begin{gather}
    \Delta\phi^2_{ent}{_{|DS}}= \frac{e^{-2r}+1/\eta -1}{8\alpha^2 e^{2r}}.
\end{gather}
Comparing Eq.~(33) with Eq.~(12), we can see that the effect of double seeding is only to enhance the phase sensitivity by a factor of 4.

Similar to this, we can also obtain the phase sensitivity corresponding to the separable configuration with a double-seeded PA as,
\begin{gather}
    \Delta\phi^2_{sep}{_{|DS}}= \frac{e^{-2r}+1/\eta -1}{8\beta^2 Me^{2r}},
\end{gather}
which again shows an improvement by a factor of 4 compared to Eq.~(17). However, the double-seeded configuration is more resource-intensive and more challenging to implement because of the PA becoming phase sensitive. 
\section{Conclusion}
In this work, we have proposed a multi-segmented optical gyroscope configuration in which unknown phase shifts are distributed at each node. Using an $M$-mode entangled bright two-mode squeezed state, we show that the average phase shift could be estimated with quantum-enhanced sensitivity. To assess the advantages obtained from the mode-entangled quantum state, we compared the sensitivity results with those obtained from the system using separate individual bTMSS. We find that both the mode-entangled and the separable configurations beat the SNL achieved by coherent states of light. And paramountly, it was observed that even with significant losses in the system, the mode-entangled configuration shows better performance than the separable configuration. This performance gain increases with the number of gyroscopes. For each particular value of the transmissivity, there exists a finite value of $N$ at which the performance gain peaks. For example, when $M=4$ and there is a loss of $5\%$ photons in each channel, the sensitivity ratio peaks at $N=2.53$ with an enhancement factor of $3.45$ dB compared to the separable configuration. 

We also derived the ultimate quantum precision bounds for both configurations and showed that, with increasing photon number, the phase sensitivity of the proposed scheme approaches the quantum Cramér-Rao bound. These results highlight the advantage of mode-entangled bTMSS for distributed phase estimation in optical gyroscopes, paving the way toward quantum-enhanced inertial navigation and precision metrology within networked sensor architectures.   

\section{Acknowledgment}
PMK \& AK acknowledge the financial support from the Indian Institute of Space Science and Technology (Govt. of India). GSA is grateful for the support of NSF Award No.2426699, the Robert A. Welch Foundation (A-1943-20240404).

\onecolumngrid
\section{APPENDIX}
\setcounter{equation}{0}
\renewcommand{\theequation}{A\arabic{equation}}
\subsection{\label{sec:level3} Mode-entangled configuration}
Since two-mode squeezed states are used for the phase estimation, we perform homodyne measurements of the superposed phase quadratures of the output modes of each gyroscope. And we collectively measure the joint phase quadrature to estimate the average phase shift. The joint phase quadrature is defined as,
\begin{gather}
    Y_+=\sum_{j=1}^M Y_j \\
    Y_j=\frac{Y_{a_j^{\prime}}+Y_{b_j^{\prime}}}{\sqrt{2}} \\
    = \frac{\cos{\phi_j}}{\sqrt{2}} \big( \sqrt{\eta} Y_{a_j} + \sqrt{1-\eta} Y_{a_{j,vac}} + \sqrt{\eta} Y_{b_j} + \sqrt{1-\eta} Y_{b_{j,vac}}\big)~~ - \nonumber \\ \frac{\sin{\phi_j}}{\sqrt{2}} \big( \sqrt{\eta} X_{a_j} + \sqrt{1-\eta} X_{a_{j,vac}} + \sqrt{\eta} X_{b_j} + \sqrt{1-\eta} X_{b_{j,vac}} \big)
\end{gather}
Further, considering the small angle approximation, the mean and variance of the joint phase quadrature become,
\begin{gather}
    \braket{Y_+}_e=-\sqrt{ \eta}\braket{X_{a_j}+X_{b_j}}\sum_{j=1}^M \sin{\phi_j} \approx -\sqrt{\eta}\alpha e^r \sqrt{M}\phi \\
    \braket{Y_+^2}_e=\frac{1}{2}\sum_{j=1}^M \cos^2{\phi_j}+\eta\big(\braket{a_1^\dag a_1}-\braket{a_1^2}+\braket{a_1^\dag b_1} -\braket{a_1 b_1} \big)(\sum_{j=1}^M \cos{\phi_j})^2 +  \nonumber \\
    \frac{1}{2}\sum_{j=1}^M \sin^2{\phi_j}+\eta\big(\braket{a_1^\dag a_1}+\braket{a_1^2}+\braket{a_1^\dag b_1} +\braket{a_1 b_1} \big)(\sum_{j=1}^M \sin{\phi_j})^2   \nonumber \\
    \braket{\Delta Y_+^2}_e=\frac{M}{2} +  \frac{M \eta }{2} (e^{-2r}-1) 
\end{gather}
Using the linear error propagation formula, the uncertainty in the average phase shift becomes, 
\begin{gather}
    \Delta\phi^2_{ent}= \frac{e^{-2r}+1/\eta -1}{2\alpha^2 e^{2r}}.
\end{gather}
The total number of photons probing the unknown phases is,
\begin{equation}
    N_{tot}=\alpha^2(\frac{e^{2r}+e^{-2r}}{2})+(\frac{e^{2r}+e^{-2r}-2}{2}).
\end{equation}
Constraining the average number of photons probing each gyroscope to be $N=\eta N_{tot}/M$, we write the Lagrangian function as,
\begin{equation}
     \mathcal{L}(r,\alpha,\lambda) = \Delta\phi^2_{ent} + \lambda(\frac{\eta N_{tot}}{M}- N) \\
    = \frac{e^{-2r}+1/\eta-1}{2\alpha^2 e^{2r}} + \lambda \Big( \frac{\eta}{2M} \Big[ (\alpha^2+1 ) ( e^{2r} + e^{-2r}) -2 \Big] -  N \Big).
\end{equation}
Solving the following equations,
\begin{gather}
    \frac{\partial \mathcal{L}}{\partial r}=0,~
     \frac{\partial \mathcal{L}}{\partial \alpha}=0,~
      \frac{\partial \mathcal{L}}{\partial \lambda}=0,
\end{gather} 
we get an implicit equation connecting $r$ and $N$,
\begin{gather}
    \eta  e^{2 r} \left( 2(M N-1)(2 e^{2 r} -e^{4 r})-6 M N+e^{8 r}+1\right)  
    -4 M N e^{4 r}-\left(e^{2 r}-1\right)^2
   \left(3 e^{2 r}+e^{6 r}-2\right) \eta^2=0.
\end{gather}
Also, for $\alpha$, we obtain the expression,
\begin{gather}
  \alpha=  \sqrt{\frac{ 2e^{2 r} (M N-\eta \cosh{2r}+\eta)}{\eta\left(e^{4 r}+1\right) }}.
\end{gather}
Solving Eq.~(A10) and (A11), one can calculate the values of $r$ and $\alpha$ for a given $N$, and substitute that in Eq.~(A6), to obtain the optimized phase sensitivity  $\Delta\phi^2_{e}$.

\subsection{\label{sec:level2}Separable bTMSS configuration}
\setcounter{equation}{0}
\renewcommand{\theequation}{B\arabic{equation}}
In the separable configuration, individual two-mode squeezed states, generated through separate parametric amplifiers (PAs), each with a coherent seed of amplitude $\beta$, are used as phase-estimating probe states. The mean and variance of the joint phase quadrature take the form, 
\begin{gather}
    \braket{Y_+}_s=-\sqrt{\eta}\braket{X_{a_j}+X_{b_j}}\sum_{j=1}^M \sin{\phi_j}\approx-\sqrt{\eta}\beta e^r M\phi \\
    \braket{Y_+^2}_s=\frac{1}{2}\sum_{j=1}^M \cos^2{\phi_j}+\eta\big(\braket{a_1^\dag a_1}-\braket{a_1^2}+\braket{a_1^\dag b_1} -\braket{a_1 b_1} \big)(\sum_{j=1}^M \cos^2{\phi_j}) +  \nonumber \\
    \frac{1}{2}\sum_{j=1}^M \sin^2{\phi_j}+\eta\big(\braket{a_1^\dag a_1}+\braket{a_1^2}+\braket{a_1^\dag b_1} +\braket{a_1 b_1} \big)(\sum_{j=1}^M \sin^2{\phi_j})    \\
    \braket{\Delta Y_+^2}_s=\frac{M}{2} +  \frac{M \eta }{2} (e^{-2r}-1)
\end{gather}
With this, the uncertainty in the average phase shift becomes,
\begin{gather}
    \Delta\phi^2_{sep}=\frac{e^{-2r}+1/\eta -1}{2\beta^2 M e^{2r}}
\end{gather}
Using Lagrangian multiplier formulation to optimize the sensitivity,
\begin{equation}
     \mathcal{L}(r,\beta,\lambda) = \Delta\phi^2_{sep} + \lambda(\frac{\eta N_{tot}}{M}- N) \\
    = \frac{\sqrt{e^{-2r}+1/\eta-1}}{2\sqrt{2}\beta M e^r} + \lambda \Big( \frac{\eta}{2} \Big[ (\alpha^2+1 ) ( e^{2r} + e^{-2r}) -2 \Big] -  N \Big).
\end{equation}
Solving the Lagrangian function as in Eq.~(A9), we get an implicit equation connecting $r$ and $N$,
\begin{gather}
    \eta  e^{2 r} \left( 2( N-1)(2 e^{2 r} -e^{4 r})-6  N+e^{8 r}+1\right)  
    -4  N e^{4 r}-\left(e^{2 r}-1\right)^2
   \left(3 e^{2 r}+e^{6 r}-2\right) \eta^2=0.
\end{gather}
Also, for $\beta$, we obtain the expression,
\begin{gather}
  \beta=  \sqrt{\frac{ 2e^{2 r} ( N-\eta \cosh{2r}+\eta)}{\eta\left(e^{4 r}+1\right) }}.
\end{gather}
Solving Eq.~(B6) and (B7), one can calculate the values of $r$ and $\beta$ for a given $N$, and substitute that in Eq.~(B4), to obtain the optimized phase sensitivity  $\Delta\phi^2_{s}$.

\subsection{Calculation of the Quantum Fisher Information Matrix}
\setcounter{equation}{0}
\renewcommand{\theequation}{C\arabic{equation}}
We use the formalism given in \cite{Nichols} to calculate the QFI matrix since the system we propose in Fig.~\ref{fig:fig2} is an $M$-mode bosonic CV system. Given an $M$-mode bosonic CV system described by its first and second statistical moments $\boldsymbol{d}_{\phi}$ and $V_{\phi}$, depending on a set of parameters $\phi=\{\phi_1,\phi_2,...\phi_M\}$, the QFI matrix elements are given by,
\begin{gather}
    F_{i,j}=tr[(\partial_{\phi_j}V_{\phi})L^{(2)}_{\phi_i}]+(\partial_{\phi_i}\boldsymbol{d}_{\phi}^T)V_{\phi}^{-1}(\partial_{\phi_j}\boldsymbol{d}_{\phi}).
\end{gather}
Here, $\boldsymbol{d}_{\phi}$ and $V_{\phi}$ are defined in Eq.~(22) and (23), respectively. $L$ is the symmetric logarithmic derivative defined by,
\begin{align}
    \hat{L}_{\phi_i}\hat{\rho}_{\phi} +\hat{\rho}_{\phi}\hat{L}_{\phi_i} =2\frac{\partial \hat{\rho}_{\phi}}{\partial \phi_i},
\end{align}
and
\begin{gather}
    L^{(2)}_{\phi_i}=\sum_{j,k=1}^M \sum_{l=0}^3 \frac{{(a_{\phi_i})}^{jk}_l}{v_j v_k-(-1)^l} {S^T}M^{jk}_l S.
\end{gather}
Here, $\{v_i\}$ are the symplectic eigenvalues of the covariance matrix $V_{\phi}$ and $S$ is the symplectic transformation that diagonalizes $V_{\phi}$,
\begin{gather}
    S V_{\phi} S^T= \bigoplus_{i=1}^{M} v_i \mathds{1},
\end{gather}
and,
\begin{gather}
   {(a_{\phi_i})}^{jk}_l =tr(S\partial_{\phi_i}V_{\phi} S^T M^{jk}_l).
\end{gather}
The set of matrices $M^{jk}_l$ have all entries zero except for the $2 \times 2$ blocks in the position $jk$ given by, 
\begin{gather}
    \{M^{jk}_l\}_{l \in \{0,1,2,3\}}=\frac{1}{\sqrt{2}}\{i\sigma_y,\sigma_z, \mathds{1}, \sigma_x \},
\end{gather}
where, $\mathds{1}, \sigma_x ,\sigma_y,\sigma_z$ are the $2 \times 2$ identity matrix and the Pauli matrices.

For calculating the symplectic transformation matrix $S$, we adopt the steps described in \cite{pirandola2009correlation}. Initially, we find the Williamson form of $V_{\phi}$, given by,
\begin{gather}
    W=Diagonal~\{v_1,v_2,...v_M\},
\end{gather}

Now, the symplectic transformation matrix is given by, $S=W^{1/2}RV_{\phi}^{-1/2}$, with $R=\Gamma U^{\dag}$. 

Here, 
\begin{gather}
    \Gamma=\frac{1}{\sqrt{2}}
\begin{pmatrix}
i & -i & 0 & \cdots & 0 \\
1 &  1 & 0 & \cdots & 0 \\
0 &  0 & \ddots & & \vdots \\
\vdots & \vdots & & i & -i \\
0 & 0 & \cdots & 1 & 1
\end{pmatrix},
\end{gather}
and $U$ is the matrix that diagonalises the $V_{\phi}^{-1/2}\Omega V_{\phi}^{-1/2}$.

 In the analysis, $M-1$ symplectic eigenvalues of $V_{\phi}$ are $1$, which will produce singular results in the QFI calculation. This is due to the presence of vacuum modes in the inputs of the SBNs in the configuration. Therefore, as in \cite{Guo}, we apply a weak thermal noise of mean photon number $10^{-6}$ to the vacuum ports of the two SBNs, which can be easily ignored in the final result. Thus by using Eqs.~(C2-C8) in Eq.~(C1), we obtain the final result given in Eq.~(28). 

\end{document}